\documentclass[epjST]{svjour}
\usepackage{graphicx}
\usepackage{color}
\pdfoutput=1
\newcommand{\F}{\mathcal{A}}
\newcommand{\p}{\partial}
\makeatletter
\newsavebox{\@brx}
\newcommand{\llangle}[1][]{\savebox{\@brx}{\(\m@th{#1\langle}\)}%
  \mathopen{\copy\@brx\kern-0.5\wd\@brx\usebox{\@brx}}}
\newcommand{\rrangle}[1][]{\savebox{\@brx}{\(\m@th{#1\rangle}\)}%
  \mathclose{\copy\@brx\kern-0.5\wd\@brx\usebox{\@brx}}}
\makeatother
\begin{document}
\title{Universality of efficiency at maximum power}
\subtitle{Macroscopic manifestation of microscopic constraints}
\author{B. Cleuren \and B. Rutten \and C. Van den Broeck}
\institute{Hasselt University, B-3590 Diepenbeek, Belgium}
\abstract{
Universal properties of efficiency at maximum power are investigated in a general setting. In particular, it is demonstrated how successive symmetries placed upon the dynamics manifest themselves at the macroscopic level. A general condition is derived for which thermodynamic devices are able to attain a reversible operation.} 
\maketitle
\section{\label{intro}Introduction}
According to Callen \cite{callen}, thermodynamic theory is rooted in the statistical properties of large systems and the symmetries of the underlying fundamental laws of physics. A profound illustration of the latter was given in 1931 by the seminal work of Lars Onsager \cite{onsager}, demonstrating that time reversibility of the microscopic dynamics becomes evident at the macroscopic level through reciprocity relations among the linear response coefficients. Experimental evidence of such relations was already known to exist in various thermoelectric systems, cf. the Seebeck and Peltier effects. With the development of stochastic thermodynamics \cite{ritort,sekimoto,jarzynski2011,seifertREV,vdbREV,introST}, such reciprocity relations were shown to exist also among the nonlinear response coefficients, see for example \cite{andrieuxJSTAT2007,astumianPRL2008}. These relations are a consequence of the fluctuation theorem, entailing a symmetry property for the probability distribution of entropy production.
\newline
Entropy production plays an important role in the performance of thermal engines. Stochastic thermodynamics provides the natural language to describe such systems. Recent work has concentrated on the efficiency of small sized thermodynamic machines, with a focus on the efficiency at maximum power (EMP) output \newline\cite{chris1,chris2,schmiedlEPL2008a,schmiedlEPL2008b,espositoPRL2009,ruttenPRB2009,seifertPRL2011,golubevaPRL2012,golubevaPRE2013,golubevaPRE2014,vandenbroeckPRE2012,vandenbroeckPRL2012,espositoPRE2012,hooyberghsJCP2013,vandenbroeckEPL2013}. Historically, EMP for heat engines was first discussed by Novikov \cite{novikov} who obtained the following result:
\begin{equation}\label{caeff}
\eta=1-\sqrt{1-\eta_c}=\frac{\eta_c}{2}+\frac{\eta_c^2}{8}+\frac{\eta_c^3}{16}\ldots
\end{equation}
with $\eta_c$ the Carnot efficiency. This result was later rediscovered by Curzon and Ahlborn \cite{curzon}. Quite remarkable, the resulting EMP is device independent and contains only the temperature ratio. This raises the question whether this efficiency can be placed at the same level as the Carnot efficiency. As was shown in \cite{chris1}, in the case of heat engines for which the heat and work flows are proportional, the first order term of Eq. \ref{caeff} is recovered. When an additional symmetry is present, also the second order term was reproduced in \cite{chris2}. However, calculations of the EMP for various model systems have revealed a large variety of EMP expressions. The purpose of this work is to investigate EMP in a model independent setting and for general energy conversion machines.  We demonstrate how symmetries and constraints at the microscopic level, combined with the fluctuation theorem, emerge at the macroscopic level via the expression for the EMP.
\section{A generic setup for energy conversion}
\begin{figure}[t!]
\centering
\includegraphics[width=2.in]{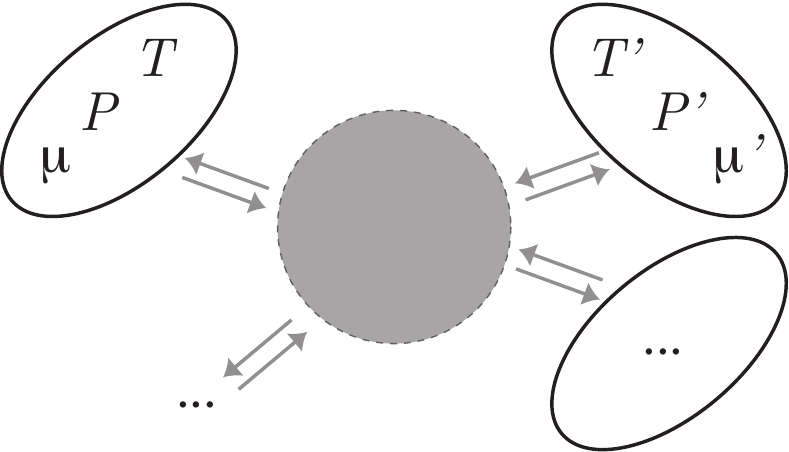}
\caption{The generic setup consisting of an arbitrary mesoscopic system in simultaneous contact with different reservoirs. Each reservoir is characterised by the values of the intensive parameters such as temperature, pressure or chemical potential.}
\label{fig:setup}
\end{figure}

Our starting point is the generic setup of a system in simultaneous contact with different reservoirs as shown in Fig.~\ref{fig:setup}. Each reservoir is characterised by the values of its intensive variables such as temperature, pressure and chemical potential. When these values are different for the various reservoirs, irreversible processes occur trying to establish a common final equilibrium state. These processes involve the exchange of extensive quantities like energy and particles for example. We consider infinitely large reservoirs so that a non equilibrium steady state is reached. For concreteness, we consider a setup which allows for the exchange of two extensive variables $X_1$ and $X_2$ when the associated thermodynamic forces (affinities) $\F_{1}$ and $\F_{2}$ are nonzero. These forces are defined as the difference of intensive variables,
\begin{equation}\label{force}
\F_{k}= F'_{k}-F_{k},
\end{equation}
which in turn are determined from the entropy fundamental relation $S(\{X_{i}\})$ as $F_{k}=\partial S/\partial X_{k}$ \cite{callen}.  Such a setup with two forces is natural in the context of thermodynamic engines where one force (e.g. a temperature difference) provides the driving energy to do work against the other force (e.g. a chemical potential difference).
\newline \noindent
The exchange processes are strongly determined by the details of the intermediate system and by its connection to the reservoirs. They are however not completely arbitrary, but must be in agreement with the microreversibility of the underlying microscopic dynamics. Such properties of the exchange processes can be quantified through the use of the cumulant generating function (CGF). Denoting $P_{t}(X_{1},X_{2})$ as the probability to have a net exchange of $X_{1}$ and $X_{2}$ for the two extensive quantities during the time interval from $0$ to $t$, the CGF is defined as
\begin{eqnarray}\label{gf}
G(\vec{\lambda};\vec{\F})&=&\lim_{t\rightarrow \infty}\frac{1}{t}\ln \langle \mathrm{e}^{-(\lambda_{1}X_{1}+\lambda_{2}X_{2})}\rangle
=\lim_{t\rightarrow \infty}\frac{1}{t}\sum_{i,j=0}^{\infty}\frac{(-1)^{i+j}}{i!j!} \llangle X_1^{i}X_2^{j} \rrangle\lambda_{1}^{i}\lambda_{2}^{j}.
\end{eqnarray}
The cumulants $\llangle X_1^{i}X_2^{j} \rrangle$ represent measurable macroscopic quantities and are determined by the specific system and its connection to the reservoirs. In particular, they depend on the intensive parameters. Replacing $F'_k$ by $F_k+\F_{k}$ and expanding in terms of the forces gives: 
\begin{equation}\label{cumexpand}
\lim_{t\rightarrow \infty}\llangle X_1^{i}X_2^{j} \rrangle/t=\sum_{k,l=0}^{\infty}L_{ij}^{kl}\F_{1}^{k}\F_{2}^{l}.
\end{equation}
The response coefficients $L_{ij}^{kl}$ are determined with respect to the equilibrium state corresponding to $F'_{k}=F_{k}$. Time reversibility imposes the following symmetry relation upon the generating function \cite{andrieux,gaspardNJP2013}
\begin{equation}
G(\vec{\lambda};\vec{\F})=G(\vec{\F}-\vec{\lambda};\vec{\F}).
\end{equation}
As a consequence, after substituting (\ref{cumexpand}) into (\ref{gf}), one can make, by rearranging the terms, the identification (see also \cite{andrieuxJSTAT2007,gaspard1}):
\begin{equation}
L_{ij}^{kl}= \sum_{m=0}^{k}\sum_{n=0}^{l}\frac{(-1)^{i+j+m+n}}{m!n!} 
L_{i+m,j+n}^{k-m,l-n}. \label{FT}
\end{equation}
This result is central to the discussion below, and expresses the macroscopic manifestation of time-reversibility at the microscopic level.\newline \noindent
For energy conversion, the focus is on the first two cumulants, i.e. the time averaged currents
\begin{equation}
J_1=\lim_{t\rightarrow \infty}\llangle X_1 \rrangle/t \;\;\mbox{and}\;\; J_2=\lim_{t\rightarrow \infty}\llangle X_2 \rrangle/t.
\end{equation}
During steady state operation these currents, together with their corresponding forces, determine the entropy production rate via the bilinear form \cite{callen}:
\begin{equation}
\dot{S}=J_1\F_1+J_2\F_2 \geq 0.
\end{equation}
In the context of thermodynamic engines, this relation can be used to obtain a general and device independent definition of efficiency. Without loss of generality we consider $J_2\F_2>0$ so that the flow $J_2$ is spontaneous and regarded here as the driving process. $\F_2$ is the corresponding driving force.  This process is used to induce a flow $J_1$ against the force $\F_{1}$, so that $J_1\F_1<0$. The efficiency of the conversion process can be defined as follows:
\begin{equation}
0 \leq \eta_{s}= \frac{-\F_{1}J_{1}}{\F_{2}J_{2}} \leq 1 \label{Seff}.
\end{equation}
This concept was coined entropic function in \cite{kedem}.\newline \noindent
\section{Microscopic constraints}
From Eq. (\ref{cumexpand}) it is straightforward to show that Onsager symmetry $L_{10}^{01}=L_{01}^{10}$ is present at the linear order in the currents. Combining Eq. (\ref{cumexpand}) with Eq. (\ref{FT}) leads to:
\begin{eqnarray}
J_{1}&=&\left(L_{20}^{00}/2\right)\F_{1}+L\F_{2}+\left(L_{20}^{10}/2\right)\F_{1}^{2}+\left[Q+L_{20}^{01}/2\right]\F_{1}\F_{2}+M\F_{2}^{2} + \ldots \label{J1}\\
J_{2}&=&L\F_{1}+\left(L_{02}^{00}/2\right)\F_{2}+Q\F_{1}^{2}+\left[L_{02}^{10}/2+M\right]\F_{1}\F_{2}+\left(L_{02}^{01}/2\right)\F_{2}^{2} + \ldots \label{J2}
\end{eqnarray}
with
\begin{equation}
L=L_{11}^{00}/2\;\;\;;\;\;\; Q=L_{11}^{10}/2 \;\;\;;\;\;\; M=L_{11}^{01}/2.
\end{equation}
More surprisingly, a similar structure is also appearing at higher orders.

\subsection{Strongly coupled flows}
In their paper \cite{kedem}, Kedem and Caplan introduced the concept "degree of coupling" between the flows $J_1$ and $J_2$. Since their work was limited to the linear order terms in Eqs. (\ref{J1}) and (\ref{J2}), the degree of coupling $q$ was defined in terms of the linear coefficients $q=L/\sqrt{L_{10}^{10}L_{01}^{01}}$. Maximal coupling is achieved for $q=\pm1$, in which case the two flows are proportional to each other.\newline \noindent
The concept of maximal coupling can be extended beyond the linear regime as follows. We call a process strongly (maximally) coupled if the exchange between reservoirs is constrained by the property:
\begin{equation}
X_{1}= \varepsilon X_{2} \label{scp}.
\end{equation}
This property will be further referred to as {\it strong coupling of the process} (SCP), and is especially relevant in nanoscale systems and models \cite{ruttenPRB2009,linkePRL2005}. Obviously it implies $J_1=\varepsilon J_2$. At the level of the response coefficients it entails:
\begin{equation}
L_{i,j}^{k,l}=\varepsilon L_{i-1,j+1}^{k,l}. \label{SCcoef}
\end{equation}
Combining this with the result from the fluctuation symmetry gives:
\begin{eqnarray}\label{stall}
J_{1}&=&L\left(\varepsilon \F_{1}+\F_{2}\right)+\left(\varepsilon \F_{1}+\F_{2}\right)\left(Q\F_{1}+M\F_{2}\right)\nonumber \\
&&\quad\quad\quad\quad\quad\quad\quad\quad+\left(\varepsilon \F_{1}+\F_{2}\right)\left(R\F_1^2+S\F_1\F_2+N\F_2^2\right) \ldots
\end{eqnarray}
One notices immediately the appearance of the combination $\varepsilon \F_{1}+\F_{2}$. A straightforward calculation, given in appendix \ref{usf}, shows that this is true at every order. Hence, one can identify a unique stalling force $\F_1=-\F_2/\varepsilon$ for which both fluxes simultaneously vanish. The efficiency is then
\begin{equation}
\eta_{s}= \frac{-\F_{1}J_{1}}{\F_{2}J_{2}}=\frac{-\varepsilon\F_{1}}{\F_{2}}=1.
\end{equation}
Hence SCP is a sufficient condition to attain reversible operation at the stalling force. Notice also that this result is independent of the specific value of $\varepsilon$.\newline \noindent
\subsection{Strongly coupled forces}
In \cite{espositoPRL2009} a stronger condition was considered. When the generalised forces only appear in the combination $\varepsilon \F_{1}+\F_{2}$, the probability distribution $P_{t, \F_{1},\F_{2}}(X_{1}, X_{2})$ can be written as $P_{t,\varepsilon \F_{1}+\F_{2}}(X_{1}, X_{2})$. We will refer to this property as \emph{strong coupling in the forces} (SCF). As consequence, the generating function has the property
\begin{equation}
\frac{\p \; G(\vec{\lambda};\vec{\F})}{\p \F_{1}}=\varepsilon \frac{\p \;  G(\vec{\lambda};\vec{\F}) }{\p  \F_{2}},
\end{equation}
which can be used on the definition of the generating function Eq. (\ref{gf}) to derive a relation similar to Eq. (\ref{SCcoef})
\begin{equation}
(k+1)L_{i,j}^{k+1,l}=\varepsilon L_{i,j}^{k,l+1}(l+1) \label{SCforces}.
\end{equation}
Using this relation, it is easy to see that we have a collapse into a single generalized force $\F=\varepsilon \F_{1}+\F_{2}$, thus also making the current expansions only dependent on $\F$, with only one free coefficient at any order. Indeed, successively applying Eq. (\ref{SCforces}) on $L_{1,0}^{p-i,i}$ yields:
\begin{equation}
	L_{1,0}^{p-i,i}=\varepsilon^{(p-i)} L_{1,0}^{0,p} {p \choose i}.
\end{equation}
Hence
\begin{equation}
J_{1}=\sum_{p=0}^{\infty} \sum_{i=0}^{p} L_{1,0}^{p-i,i}\F_{1}^{p-i}\F_{2}^{i}=\sum_{p=0}^{\infty} L_{1,0}^{0,p} (\varepsilon \F_{1}+\F_{2})^{p}.
\end{equation}
For the determination of the second current $J_2$, we show in Appendix B that the combination of SCF and the fluctuation symmetry automatically implies SCP, that is $J_1=\varepsilon J_2$. The proportionality constant between the currents $\varepsilon$ is the same as in $\varepsilon \F_{1}+\F_{2}$:
\begin{eqnarray}\label{jscf}
J_{1}&=& L \F+M\F^{2} + N\F^{3}+O(\F^{4});\\
J_{2}&=&J_1/\varepsilon.
\end{eqnarray}
\section{Implications on the efficiency at maximal power}
It is clear from the previous discussion that the number of kinetic coefficients is reduced when microscopic constraints are imposed upon the system. We now clarify how this reduction of coefficients is reflected in the efficiency at maximum power. The maximum is determined in the following way: given the driving force $\F_2$, what value of the loading force $\F_1$ maximises the power output $-\F_{1}J_{1}$? The answer requires solving the following nonlinear equation:
\begin{equation}
\frac{\partial}{\partial \F_1}\left(-\F_{1}J_{1}\right)=-J_{1}-\F_{1}\frac{\partial J_{1}}{\partial \F_1}=0.
\end{equation}
Other optimisation schemes (see for example \cite{seifertREV,kedem}) are not considered here.\\
Since an analytical solution is not possible in general, we resort to a series solution
\begin{equation}\label{coeff}
\F_1=c_1\F_2+c_2\F_2^2+c_3\F_2^3+\ldots
\end{equation}
and determine the coefficients.
We already gave the current in its most general form Eqs. (\ref{J1}-\ref{J2}), due to the FT. Solving for maximal output and using the coupling parameter $q$, the EMP to first order in $\F_{2}$ reads
\begin{equation}
\eta=\frac{q^2}{4 - 2 q^2}+
\frac{q\left[\begin{array}{c}
\sqrt{L_{20}^{00}L_{02}^{00}}\left(q^2(4+q^2)L_{20}^{10}L_{02}^{00}+4L_{20}^{00}\left(8M+q^2L_{02}^{10}\right)\right)\\
-4L_{20}^{00}q\left(\left(4+q^2\right)Q+2L_{20}^{01}\right)L_{02}^{00}-8q(L_{20}^{00})^2 L_{02}^{01}
\end{array}\right]}
{8(L_{20}^{00})^2 L_{02}^{00}(q^2-2)^2}\F_2+\ldots
\end{equation}    
For a system with SCP, the coefficients from Eq. (\ref{coeff}) appear directly in the efficiency:
\begin{equation}\label{escc}
\eta=-\varepsilon \F_1/\F_2=-\varepsilon \left(c_1+c_2\F_2+c_3\F_2^2+\ldots \right).
\end{equation}
The efficiency expansion substantially simplifies:
\begin{equation}\label{emp_scp}
\eta=\frac{1}{2}-\frac{Q\F_2}{8L\varepsilon}-\frac{Q^2-2LR-2MQ\varepsilon+2LS\varepsilon}{16L^2\varepsilon^2}\F_2^2+\ldots
\end{equation}
From which we already see at lowest order the known universal factor $\frac{1}{2}$ \cite{chris1}. The latter expansion is especially relevant for autonomous isothermal motors since the same definition of efficiency (\ref{Seff}) applies. The expansion simplifies further when the forces are strongly coupled (\ref{SCforces}). From Eq. (\ref{emp_scp}) and substituting $Q=\varepsilon M$, $R=\varepsilon^2N$ and $S=2\varepsilon N$ we find:
\begin{equation}\label{emp_scf}
\eta_{s}=\frac{1}{2} - \frac{M}{8L}\F_{2}+\frac{M^{2}-2LN}{16L^{2}}\F_{2}^{2}+O(\F_{2}^{3}).
\end{equation}
\section{Discussion and perspectives}
The transition of the 18th to the 19th century marks the beginning of the industrial revolution. A prominent ingredient was the development of efficient steam engines. James Watt and his colleague engineers succeeded in improving the performance by a factor of twenty. In the wake of these developments, Sadi Carnot published a manuscript entitled "Sur la puissance motrice du feu". He showed that the efficiency of thermal machines is bounded by that of a machine operating reversibly. This insight will lead, with the notable input of Clayperon, Clausius and Kelvin, to the development of thermodynamics. Clayperon drew the attention to the fundamental significance of the Carnot cycle. Clausius introduced the state function entropy and formulated the second law of thermodynamics. The name of Kelvin is attached to the scientifically well defined concept of temperature.

In its original formulation, thermodynamics describes macroscopic systems at equilibrium. With the work of Onsager and Prigogine, it was extended to describe systems in local equilibrium. More recently, one has been able to formulate thermodynamics for small system far from equilibrium. Stochastic thermodynamics is arguably the simplest such formalism.
The most spectacular advance are probably the integral and detailed fluctuation theorems, which replace the  positivity of the entropy production by a symmetry property for its probability distribution. This result is valid for any size of the system and for any nonequilibrium state.
Revisiting the question of efficiency of thermal and other machines from this point of view has led to some remarkable discoveries. Onsager symmetry implies that efficiency at maximum power with respect to the load in the regime of linear irreversible thermodynamics is at most half of the reversible efficiency \cite{chris1}, cf. Eq. (\ref{escc}). The constraint, imposed by the fluctuation theorem, has also an impact on the nonlinear response coefficients \cite{gaspard1}. When asking the right question about efficiency, it leads to universal values for the quadratic nonlinearity for strong coupling and an appropriate additional property. In the present paper, we have shown that strong coupling with fluctuation symmetry implies an expansion of the form given in Eq. (\ref{jscf}). The additional requirement to fix the value of $M$ depends both on the system under consideration and on the definition of the efficiency. For example, for appropriate systems with "left/right" symmetry, the coefficient $M$ becomes zero \cite{chris2}. For appropriate thermal machines with the efficiency "defined a la Carnot", the coefficient is equal to $1/8$ \cite{espositoPRL2009}. These universal values can be seen as manifestations of micro-reversibility (Liouville's theorem or unitary evolution).

We finally mention another extremely exciting recent development. The efficiency of small scale systems is fluctuating. The fluctuation theorem has implications on the probability distribution for this so-called stochastic efficiency \cite{naturecommunications,gingrich2014,prx,polettiniPRL2015,proesmansEPL2015,proesmansNJP2015}. The most surprising result is that the reversible efficiency becomes the least probable in the long time limit for time-symmetric driving \cite{naturecommunications,proesmansEPL2015}. For time-asymmetric driving the probability distributions (or more precisely the large deviation functions) for forward and backward driving cross at the reversible efficiency \cite{gingrich2014,prx}. Hence the reversible efficiency is a special point for stochastic efficiency. It remains to be seen whether this result has practical implications, with in particular the absolute measurement of the temperature. The situation is reminiscent of the Jarzynski equality, which allows the measurement of equilibrium free energy differences by a large number of nonequilibrium experiments instead of a single reversible one. Here the reversible measurement of Carnot efficiency is replaced by multiple irreversible measurements from which the least likely efficiency can be identified.
\appendix
\section{Unique stalling force}\label{usf}
Using the fluctuation relation Eq. (\ref{FT}) and the SCP Eq. (\ref{SCcoef}), we will now give the proof of Eq.(\ref{stall}). The first step is to show that the following quantity 
\begin{equation}\label{lambda}
\Lambda (p)=\sum_{i=0}^{p} (- \varepsilon)^{i} L_{1,0}^{p-i,i}=0.
\end{equation}
Applying the fluctuation theorem Eq. (\ref{FT}) on the coefficients $L_{1,0}^{p-i,i}$ yields:
\begin{equation}
\Lambda(p)=\sum_{i=0}^{p} (- \varepsilon)^{i}  \left[ \sum_{m=0}^{p-i} \sum_{n=0}^{i} \frac{(-1)^{1+m+n}}{m!n!}L_{m+1,n}^{p-i-m,i-n}\right]  \label{algorde1}.
\end{equation}
Repeatedly using SCP Eq. (\ref{SCcoef}) gives after $n$ times:
\begin{eqnarray}
\Lambda (p)
&=&\sum_{i=0}^{p} (- \varepsilon)^{i}  \left[ \sum_{m=0}^{p-i} \sum_{n=0}^{i} \frac{(-1)^{1+m+n}}{m!n!}L_{m+1+n,0}^{p-i-m,i-n} \frac{1}{\varepsilon^{n}} \right] \nonumber \\ 
&=&\sum_{i=0}^{p} (- \varepsilon)^{i}  \left[  \sum_{n=0}^{i} \frac{1}{n! \varepsilon^{n}}  \sum_{m'=n}^{p-i+n}\frac{(-1)^{1+m'}}{(m'-n)!}L_{m'+1,0}^{p-i-(m'-n),i-n}\right].
\end{eqnarray}
In the second step the substitution $m'=m+n$ was used. Next we substitute the index $n=0...i$ by $l=i-n=i...0$.
\begin{equation}
\Lambda(p)=\sum_{i=0}^{p} (- \varepsilon)^{i}  \left[  \sum_{l=0}^{i} \frac{1}{(i-l)! \varepsilon^{i-l}}  \sum_{m'=i-l}^{p-l}\frac{(-1)^{1+m'}}{(m'-(i-l))!}L_{m'+1,0}^{p-l-m',l}\right].
\end{equation}
Now we simplify with the following substitution $m''=m'+l$. The double primes are dropped for ease of notation.
\begin{equation}
\Lambda(p)=\sum_{i=0}^{p} (- \varepsilon)^{i}  \left[  \sum_{l=0}^{i} \frac{1}{(i-l)! \varepsilon^{i-l}}  \sum_{m=i}^{p}\frac{(-1)^{1+m-l}}{(m-i)!}L_{m+1-l,0}^{p-m,l}\right].
\end{equation}
Next we take the coefficient $L_{i,j}^{k,l}$ outside the summation over $i$. This requires a double switch of the index $i$. First the sums over $i=0...p$ and $m=i...p$ can be replaced by the sums $m=0...p$ and $i=0...m$ where the order of summation has changed:
\begin{equation}
\Lambda(p)=\sum_{m=0}^{p} \sum_{i=0}^{m}\sum_{l=0}^{i} \frac{(-1)^{i}\varepsilon^{l}}{(i-l)!} \frac{(-1)^{1+m-l}}{(m-i)!}L_{1+m-l,0}^{p-m,l}.
\end{equation}
Second, a similar switching is done from $i=0...m$ and $l=0...i$ to $i=l...m$ and $l=0...m$ which makes
\begin{equation}
\Lambda(p)=\sum_{m=0}^{p} \sum_{l=0}^{m} \varepsilon^{l}L_{1+m-l,0}^{p-m,l} (-1)^{1+m-l} \left[ \sum_{i=l}^{m} \frac{(-1)^{i}}{(i-l)!(m-i)!}\right].
\end{equation}
The sum appearing between square brackets yields $(-1)^{l}\delta_{l,m}$ and hence
\begin{equation}
\Lambda (p)=\sum_{m=0}^{p} \varepsilon^{m}L_{1,0}^{p-m,m}(-1)^{m+1}=-\Lambda (p)=0.
\end{equation}
It can be used to find the general stalling force of the process. Using Eq. (\ref{lambda}) on the first coefficient $L_{1,0}^{p,0}$ of the $p$-th order in the current expansion of $J_{1}$ gives
\begin{equation}
J_{1}^{(p)}= \sum_{i=0}^{p} L_{1,0}^{p-i,i} \F_ {1}^{p-i}\F_{2}^{i} \label{ordep}=\sum_{i=1}^{p}L_{1,0}^{p-i,i} \F_{1}^{p-i}(\F_{2}^{i}- (-\varepsilon)^{i} \F_{1}^{i}).
\end{equation}
Finally, since $x^{i}-y^{i}=(x-y)(\sum_{j=0}^{i-1}x^{i-1-j}y^{j})$, this result proofs that every order $p$ contains the factor $\varepsilon \F_{1}+\F_{2}$.
\section{SCF implies SCP}
We now show that strong coupling in the forces implies also strong coupling in the process. We will proof that when the thermodynamic forces $\F_{i}$ are strongly coupled Eq. (\ref{SCforces}), it automatically implies strong coupling in the process Eq. (\ref{SCcoef}). In the main text it was already established that the strong coupling relation Eq. (\ref{SCforces}) allows us to write the expansion of the cumulant $i,j$ as follows:
\begin{equation}
\llangle X_{1}^{i}X_{2}^{j}\rrangle=t \sum_{p=0}^{\infty} L_{i,j}^{0,p}\F^{p}
\end{equation}
with $\F=\varepsilon \F_{1}+\F_{2}$. So to prove the strong coupling in the process, the relation
\begin{equation}
L_{i+1,j}^{0,p}=\varepsilon L_{i,j+1}^{0,p}
\end{equation}
must hold.
From the strong coupling in the forces Eq. (\ref{SCforces}) it follows:
\begin{equation}
L_{i,j}^{1,p}=\varepsilon (p+1) L_{i,j}^{0,p+1}.
\end{equation}
Using the fluctuation relation Eq. (\ref{FT}) on the lhs leads to
\begin{eqnarray}
L_{i,j}^{1,p}&=& \sum_{m=0}^{1}\sum_{n=0}^{p} \frac{(-1)^{(i+j+m+n)}}{m!\;n!} L_{i+m,j+n}^{1-m,p-n}\nonumber \\
&=&\sum_{n=0}^{p} \frac{(-1)^{i+j+n}}{n!} L_{i,j+n}^{1,p-n} + \sum_{n=0}^{p} \frac{(-1)^{i+j+1+n}}{n!} L_{i+1,j+n}^{0,p-n} \label{tussen}
\end{eqnarray}
The last term can be rewritten with the fluctuation theorem Eq. (\ref{FT}):
\begin{equation}
L_{i+1,j}^{0,p}=\sum_{n=0}^{p} \frac{(-1)^{i+1+j+n}}{n!} L_{i+1,j+n}^{0,p-n}.
\end{equation}
So Eq. (\ref{tussen}) can be written as
\begin{equation}
L_{i,j}^{1,p}=\sum_{n=0}^{p}\frac{(-1)^{i+j+n}}{n!}L_{i,j+n}^{1,p-n}+L_{i+1,j}^{0,p}.
\end{equation}
Rearranging the terms then leads to
\begin{equation}
L_{i+1,j}^{0,p}=\varepsilon \left[ (p+1)L_{i,j}^{0,p+1}- \sum_{n=0}^{p}\frac{(-1)^{i+j+n}}{n!}(p-n+1)L_{i,j+n}^{0,p-n+1} \right]
\end{equation}
where Eq. (\ref{SCforces}) was used on the coefficients $L_{i,j+n}^{1,p-n}$ of the last term. 
Using the FT Eq. (\ref{FT}) then on $L_{i,j}^{0,p+1}$:
\begin{equation}
L_{i+1,j}^{0,p}=\varepsilon \left[ (p+1)\sum_{n=0}^{p+1}\frac{(-1)^{i+j+n}}{n!}L_{i,j+n}^{0,p+1-n}- \sum_{n=0}^{p}\frac{(-1)^{i+j+n}}{n!}(p-n+1)L_{i,j+n}^{0,p-n+1} \right].
\end{equation}
The two sums can be merged together:
\begin{equation}
L_{i+1,j}^{0,p}=\varepsilon \sum_{n=0}^{p+1}\frac{(-1)^{i+j+n}}{n!} (n)L_{i,j+n}^{0,p-n+1}.
\end{equation}
The term $n=0$ can be dropped and substituting the index $m=n-1$ yields the desired result.
\begin{equation}
L_{i+1,j}^{0,p}
=\varepsilon \left[ \sum_{m=0}^{p}\frac{(-1)^{i+j+m+1}}{m!} L_{i,j+m+1}^{0,p-m}\right]=\varepsilon L_{i,j+1}^{0,p}
\end{equation}
where in the last step the FT Eq. (\ref{FT}) was used again. So the strong coupling in the forces has been shown to imply the strong coupling of the process Eq. (\ref{SCcoef}).


%

\end{document}